\documentclass[prE,twocolumn,showpacs]{revtex4}
\usepackage[T1]{fontenc}
\usepackage[cp1250]{inputenc}
\usepackage{color}
\usepackage{graphicx}
\usepackage{hyperref}

\begin{document}

\title{Dynamics of hate based networks}

\author{Pawel Sobkowicz}
\affiliation{Warsaw, Poland}
\email{pawelsobko@gmail.com}
\homepage{http://countryofblindfolded.blogspot.com}

\author{Antoni Sobkowicz}
\affiliation{Faculty of Physics, Warsaw University, Hoża 69, PL-00-681 Warsaw, Poland}

\begin{abstract}
We present a study of the properties of network of political discussions
on one of the most popular Polish Internet forums. This provides the opportunity to study the 
computer mediated human interactions
in strongly bipolar environment. The comments of the participants
are found to be mostly disagreements, with strong percentage of invective
and provocative ones. Binary exchanges (quarrels) play significant role
in the network growth and topology. Statistical analysis shows that the growth of
the discussions depends on the degree of controversy of the subject  and
the intensity of personal conflict between the participants. This is in
contrast to most previously  studied social networks, for example networks of scientific
citations, where the nature of the links is much more positive and
based on similarity and collaboration rather than opposition and abuse. 
The work discusses also 
the implications of the findings for more general studies of consensus
formation, where our observations of increased conflict 
contradict the usual assumptions that interactions 
between people lead to averaging of opinions and agreement.
\end{abstract}

\pacs{89.75.-s, 89.75.Da, 89.75.Hc}

\maketitle

\section{Introduction}

Most of the studies of social networks concentrate on properties of
groups formed due to attraction among participants. 
In such situations the links form between actors sharing
some similarity, for example common interest (in the case of scientific collaboration 
networks or cross-linked Internet sites) or likeness of views 
(in the case of political associations). The networks of
social interactions grow via preferential attachment on a `rich get richer' principle.
 In most
situations, differences and conflicting  positions and opinions are viewed as
limitations and barriers to such network formation. Frequent psychological reaction to
meeting with someone who holds an opposing view is not an attempt to
convince (phenomenon assumed widely in the {}``consensus formation''
models), but rather cutting off the connection. In face-to-face encounters
this avoidance limits growth of networks based on contrariness. Perhaps the
most developed form of such direct `hate based networks' might be long term
family or tribal feuds. The advance of modern technologies has, however,
provided an indirect contact field, where it is possible to express
contrary views, hate-filled reactions and aggressive attacks without
the risk of physical injury or personal danger. This `bravery of being
out of range' allows such networks to form and flourish. In this paper
we present a study of specific networks that benefit and grow thanks
to disagreements and hate. 

A good example of such networks is offered by user comments to news
items published on the Internet -- feature common to most of the web portals today. 
The ease of access and anonymity that the Internet has provided greatly
enhanced the number of people participating in such discussions. From the research point of view,
such discussions are relatively easy to document and can provide 
necessary data for meaningful statistical work.

For our study we have chosen discussion forums at one of the most
popular Internet portals and news sites in Poland, \url{http://www.gazeta.pl}.
Specifically, we have limited our research to discussions spurred by
the \emph{Politics} subsection of the news. Current political situation in
Poland makes it an almost ideal ground for such a study. There
is  almost clearly bipolar split between the two main
political factions (Platforma Obywatelska, PO, and Prawo i Sprawiedliwo\'{s}\'{c},
PiS). The conflict shown at the highest positions of the state 
is even more visible in the group of active readers of Internet portals.
 In fact the participants
are probably better characterized as belonging to the anti-PO and
anti-PiS groups rather than the `pro' counterparts.

The  reason for choosing this particular forum is the fact that
while preserving anonymity of the users, it also provides relative
recognizability of participants. This results from the fact that
only registered users are allowed to post comments,
and can be identified by registered nicknames. 
We may assume that participant XYZ in one discussion thread
is the same person as participant XYZ in a different one. 
Of course this leaves open the possibility of a single real person
using multiple Internet personalities. However, even with this limited
trackability, it is possible to try to find hubs of communication,
both in the comment writing  and in reaction
to published comments.

Our goal is to find if the change in motivations for linking 
from positive to negative changes the general network properties.

\section{Methods }

The data for the study have been gathered using a dedicated program, 
written for the purpose of loading and initial analysis
of the discussion threads at the selected site. The program performed
automated tasks of data collection and cleaning and enabled the next step, which
consisted of assigning political stance to discussion participants
and to classification of the comments. This part of the analysis, by far most time
consuming and cumbersome, had to be done by a human, by reading all the comments in a thread.

It should be noted that in almost all cases (with a single exception
only), the whole discussions were actually linked not to the original
news article but to the first comment. This is a result of operational
process of the portal, and the fact that pushing the `comment' button
in typical situations links the post not to the original source
but to the earliest existing post. To avoid spurious statistics
this phenomenon has been corrected for by the program.

The participants were assigned three possible types (called \emph{nodeclass}).
The first was for commentators whose viewpoints were visibly in agreement with
one of the two factions (nodeclassses A and B).
The remaining participants,
for whom it was impossible to clearly assign position were given nodeclass
NA.

The comments were  classified according to the following scheme:
\begin{description}
\item [{Agr}] - comment agrees with the covered material (either the original
news coverage or the preceding comment in a thread);
\item [{Dis}] - comment disagrees with the covered material (either the
original news coverage or the preceding comment in a thread);
\item [{Inv}] - comment is a direct invective and personal abuse of the previous
commentator;
\item [{Prv}] - provocation - comment is aimed at causing dissent, often
only weakly related to the topic of discussion;
\item [{Neu}] - comment is neutral in nature, neither in obvious agreement
or disagreement;
\item [{Jst}] - `just stupid' comment, which is totally unrelated to the
topic of discussion;
\item [{Swi}] - comment signifying a switch in participant's position leading to agreement 
between two previously opposing commentators.
\end{description}
Other works on computer mediated discussions in closed communities
used different message classification themes, for example 
\citet{jeong03-1,jeong05-1} has proposed grouping comments
by categories such as `Arg'  -- argument for a given thesis (corresponding
to our Agr category); `But' -- a challenge (corresponding to our Dis category),
`Expl' -- for posts giving explanations, and `Evid' for posts giving factual
evidence. In our case, the explanations
and evidence posts were rather scarce, possibly due
to political nature of the disputes. We have therefore opted for categorization
that reflected the emotional nature of communication, rather than
factual one. 

Following the process of categorization we performed
standard analyses typical for network systems. As the literature on
the subject is very rich we refer here to the general overviews,
for example \citet{dorogovtsev03-2,albert02-1,newman00-1,newman02-1}.
It should be noted
that the average size of the network formed by  
posts related to a single news item was relatively small (from a
few tens to a few hundreds of comments, thus the statistical spread
of results for single discussion threads was rather significant.

In our analysis we have used publicly available program GUESS, developed
and maintained by Eytan Adar (\url{http://graphexploration.cond.org/index.html},
see also \citet{adar06-1,adar07-1}). 

To understand the data we have developed a computer simulation model, which has resulted in quantitatively comparable system characteristics, allowing to understand the role of the most important factors driving the growth of comment networks. Details of the model are given in Section~\ref{SecSimul}.
{The programs and scripts used in analysis are available from the authors on request.}

%--------------------------------------------
\section{Results}

\subsection{General statistics of discussions and temporal dynamics }

The statistical properties of discussion threads depend, obviously, 
on the visibility of the news stories they relate to. 
Some of the news are featured on the portal opening page, 
so one would expect that this should influence the number 
of user comments. Our observations do not confirm these expectations -- 
the advantage of the `front page news' is not significant. 
The users activity does not follow editors choices.
Within the in the \emph{Politics} category each day, 
the portal carries  between 5 and 20 news items each day. 
On a typical graphics display, a 
visitor sees  4--6 most recent news items (although the web page has also 
a `most commented' section, allowing a short cut to older, but popular stories). 
While the `screen space' and graphical clues give no preferences 
(order of presentation is strictly temporal), 
the number of comments spurred by each story varies significantly, depending on its content.

%Political 
Discussion size distribution shows a definite fat-tail behaviour. In addition to  news items that raise no commentary at all, and weakly commented ones (below 50-70 comments), there are quite a few mid-sized discussions (up to about 200 comments) and occasional extended discussions (between 200 and 500 posts).

The news-related discussions are, by their very nature, short-lived. While the portal allows to view and comment 
on news backdating more than 2 weeks, the comment frequency vanishes rapidly with time. Usually, there are very 
few comments later than 24 hours after publication, and practically none after 48 hours. In fact numerical 
analysis of the threads shows for many discussions a reasonably good fit with exponential decay timeline, with 
half-life of between 1 and 4 hours. There are exceptions to this, for example news which gain popularity many 
hours after publication (this happens usually for stories published at night and commented during business 
hours) or stories which get a `second life' due to a quarrel between a few participants.

\subsection{User and comment statistics}

Our first task was to study the typical properties of the network
of comments, namely the activity of users measured
by their indegree and outdegree statistics. 
It should be stressed here that network structure grows by addition of unique post-to-post links. 
Therefore translating the post-to-post network to a 
user-to-user one, by necessity, introduces multiple connections between users. 
In our analysis we treat these connections as 
separate, which is reflected in several statistical properties.
Outdegree $k_o$ of a user, which
corresponds to a number of comments a given he or she posts in a discussion
(or cumulatively in many discussions), measures the `productivity'. We have attempted to fit the 
outdegree distributions for mid-sized and large discussions (where such measurements can be meaningful) 
with a modified power law  
$P(k_{o}) = A (k_{o}+c_o)^{-\alpha}+H(k_{o,max}-k_o)$. 
The first term has been used previously by \citet{newman01-1} 
in their analysis of the connectivity of internet sites.
The additional step function $H(k_{o,max}-k_o)$ reflects existence of several users with exceptionally 
high number of posts; an explanation of  their origin shall be presented in later part of the paper.
It is interesting to note that  for most discussions
the value of the $c_o$ constant is rather small ($|c_o|<1$).

We turn next to user indegree, $k_i$. This measures
not the author's activity, but the response to his or her posts, in
some way related to their `interest' value, or to the amount of controversy they raise. 
We note that
there are quite a few posts that do not elicit any comment, i.e. with
indegree equal to zero.  Here also we obtained
a reasonably good fit by using  modified power law $P(k_{i}) = A (k_{i}+c_i)^{-\alpha}+H(k_{i,max}-k_i)$; 
 however,  the value of $c_i$ is much larger. 
The power law exponent $\alpha$ for indegree is significantly larger than for the outdegree, 
indicating a much faster drop of the typical popularity than of productivity.

Interestingly, in almost all mid-sized and extended
discussions we find small groups of participants with unusually high indegree and outdegree
values, much above the predictions of the power-law. These users are responsible for 
the need for additive step function part in the distribution. 
The explanation of their origin
is simplified by observation that they are in many cases the same users. High numbers
of the connections result not from the random
process of preferential attachment (typical for internet sites or scientific citations), 
but from extended  exchanges of posts between pairs of users.
Needless to say, most of such exchanges are  confrontational, filled with disagreements and abuse. 
For this reason we'll use the term `quarrels' to describe them.
Moreover, such verbal duels are easily visible in the 
graphical view of the comments web page. Because of this visibility, they attract 
additional comments from supporters of each of the quarrelling sides. 
Quarrels increase $k_o$ and $k_i$ of their participants and thus  change the general
degree distributions. 
Typically  quarrels longer 5-7 exchanges take only between 3 and 7 
percent of the total number of comments, but we have observed discussion 
threads where such the ratio was much higher, for example 21\% of the 
220 posts in a thread resulted from just two long exchanges.

Recurrence of user nicknames connected with quarrels in various threads has added plausibility to a 
hypothesis that they  are largely due to the presence of `duellists' -- users seeking each other's comments
and joining in the fights. For such users the growth of $k_i$ and $k_o$ should be correlated. 
To test these ideas, we have performed cumulative statistical analysis
for all participants in a set of 58 discussions. This has been done
using assumption that the identity of real participants remains fixed
to nicknames within the whole scope of the portal. Results are quite interesting. 
Out of almost 2000 users there were only a few with high values of indegree (16 with $k_i\ge50$).
Similarly, there were only 23 users with $k_o\ge50$.
Eleven users belonged to both these groups. 
The average outdegree was $\left\langle k_o\right\rangle = 4.65$, 
while indegree (excluding references to the original news items to 
count only post-to-post links)  was
$\left\langle k_i\right\rangle = 3.04$.

In addition to duellists in the studied discussions we have found a  group of hyperactive users specializing in
abusive comments (known as \emph{trolls}) who, while publishing a lot of 
comments, receive much smaller number of replies. For example
one user has posted 236 times receiving only 51 replies.
Although trolls post highly provocative comments they are frequently ignored  -- 
most users seem to know the rule `\emph{don't feed the troll}'.

\begin{figure*}
\includegraphics[scale=0.8]{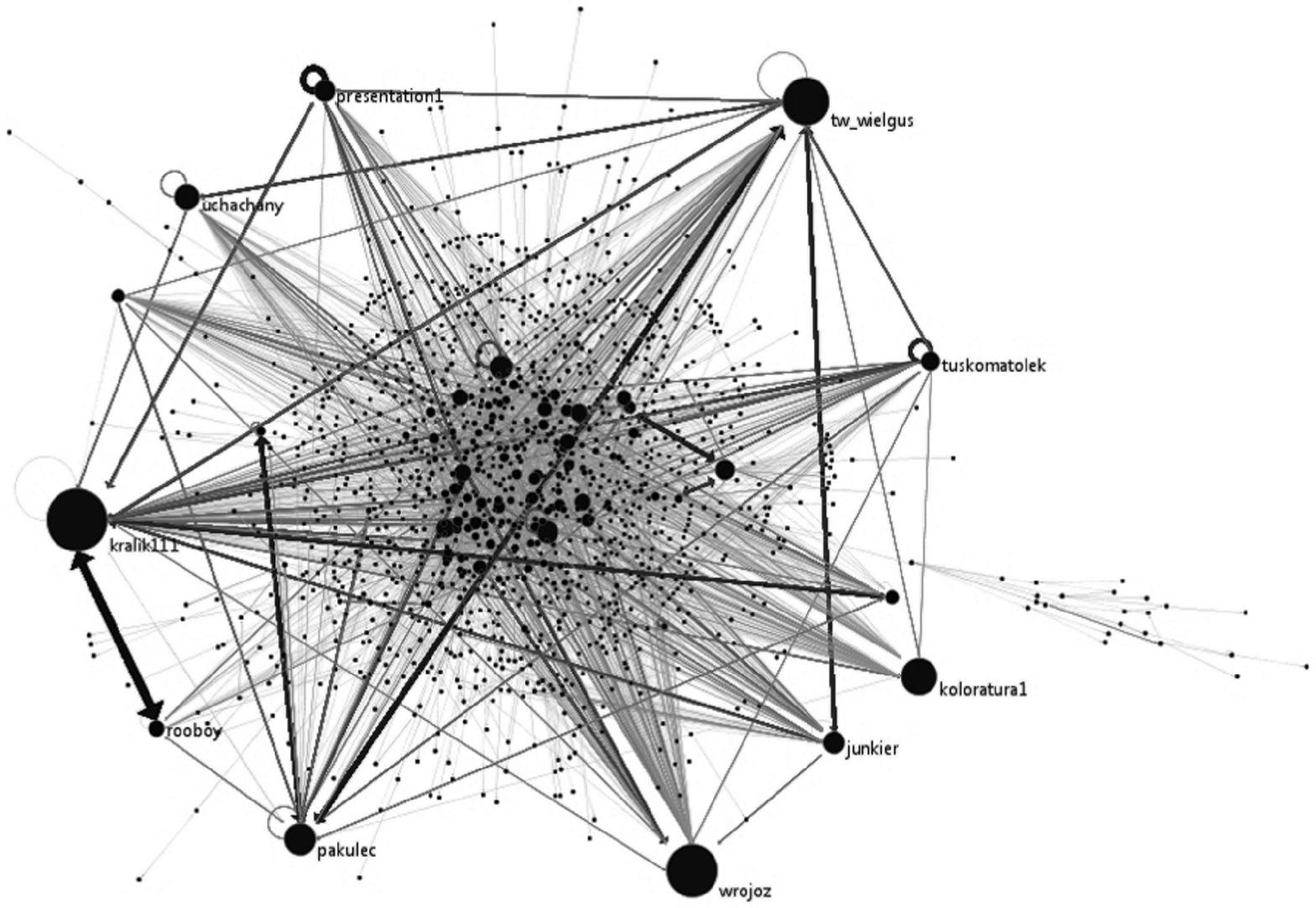} 
\includegraphics[scale=0.8]{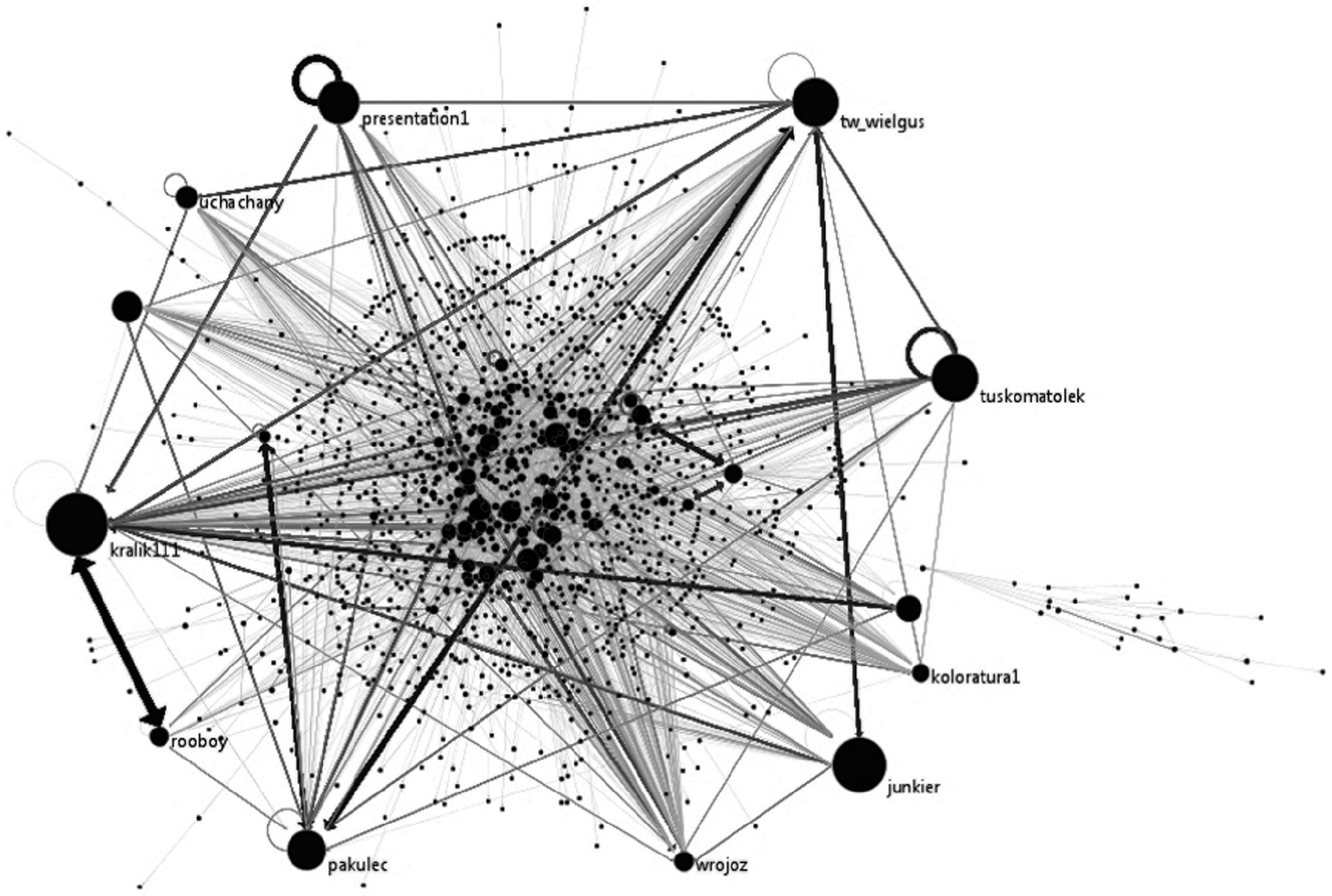} 
\caption{Two views of the topology of the network connecting the users participating in 58 large and mid-size discussions within one month. Top panel: size of the nodes corresponds to outdegree of the user. Bottom panel: size of the node corresponds to indegree. 
Links width  reflects the number of communications between the users 
-- binary exchanges are clearly identifiable. Some users have been identified by their nicknames. 
This allows to identify notorious \emph{trolls} (such as \texttt{wrojoz} and \texttt{koloratura1}), 
who have many posts but relatively few responses, and controversy leaders, 
such as \texttt{tuskomatolek} and \texttt{junkier} (who have more responses that the posts). 
Despite the fact that almost 2000 users have participated in the discussions, 
only a few of them dominate the exchanges,  by their posting activity and 
by the concentration of responses, such as \texttt{kralik111}. 
A perfect example of a user whose participation in discussions 
is motivated by negation and abuse of a particular opponent is given by \texttt{rooboy}.\label{FigTopology}}
\end{figure*}

\begin{figure*}
\includegraphics[scale=0.75]{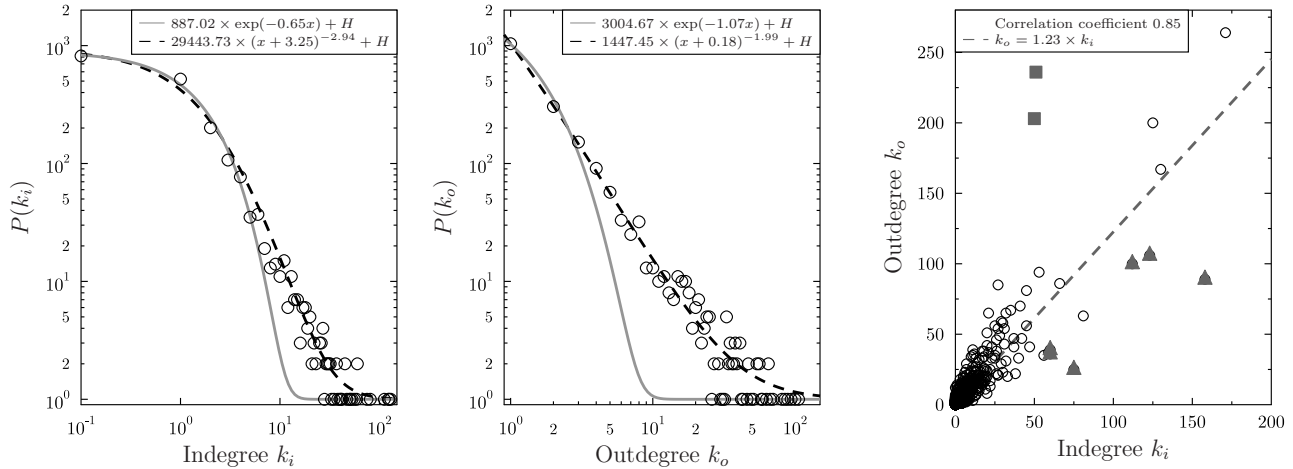} 
\caption{Cumulative indegree and outdegree distributions for  58 Politics mid-size and large discussions over a period of 30 days. The third panel shows correlation between $k_i$ and $k_o$, with two squares indicating the notorious trolls, i.e. individuals posting a lot of comments but getting only a few answers. The triangles indicate the most active -- and at the same time popular users, who could be called `controversy leaders', receiving significantly more comments than they post. To be able to show the posts that have not resulted in any comment (with indegree equal zero) on the log-log scales we have artificially shifted them to $k_i=0.1$. \label{FigCumulative}  
}
\end{figure*}

Figure~\ref{FigTopology} shows the cumulative network topology for the 58 analysed discussions. There were 1977 users, 9135 posts, out of which 3194 were linked directly to news items. In the figures we have removed all such 
links, leaving only connections between the users. The two views focus on outdegree (upper panel) and indegree. Multiple connections between pairs of users are colour and width coded, to emphasize binary exchanges. We can clearly see how a few of the users seem to dominate the whole forum. Figure~\ref{FigCumulative} presents the cumulative distributions $P(k_i)$ and $P(k_o)$, as well as correlation between $k_i$ and $k_o$ values. The two quantities are highly correlated, with the correlation coefficient of 0.85.

From the texts of the comments we know that in many cases the motivation for posting was a drive to achieve popularity (or at least notoriety). It is interesting to compare this notion with the observations of
\citet{huberman08-1}, who identified the drive to achieve fame and
visibility as one of the distinct factors determining the number of
posts on \texttt{YouTube}. The authors have shown that `\emph{the productivity
exhibited in crowdsourcing exhibits a strong positive dependence on
attention. Conversely, a lack of attention leads to a decrease in
the number of videos uploaded and the consequent drop in productivity,
which in many cases asymptotes to no uploads whatsoever.}' This  is
supported in our case by the fact that the identities of the most productive and
most commented on participants, summed over the set of threads are
highly correlated (third panel in Figure~\ref{FigCumulative}). We observe a crowd of `one-comment' participants
and a few popular and prolific ones. Moreover, the \emph{duellists} recognize each other and 
tend to join in the sub-threads with remarks of `\emph{Oh, it's you again, you ***!}', 
which, of course, spur new rounds of abuse. 

An interesting psychological observation is the existence of impersonators
of `famous' commentators. They choose a nickname that is on the first glance 
identical to  the original one,  for example by adding unobtrusive parts to the user name,
such as changing from XYZ to XYZ.
-- which often goes unnoticed. This is the most aggressive form of \emph{trolling}.
In most cases  the views of the original user and the impostor are radically different. 
The troll's intention is to create chaos and confusion, as an unsuspecting reader often finds 
comments with radically different views or even exchanges between apparently the same participant, quarrelling with himself.

\subsection{Comment classification}
In addition to rather simple  counting of links between posts and users 
we wanted to goal was to study the \emph{content} and \emph{tone} of these comments. 
Our intention was to check if the forum provided any chance of reaching a consensus, 
or even of decreasing the level of disagreement. This part of the evaluation required 
human evaluation, and the results of assignment of each comment to a given class 
are obviously less objective. In some cases we admit to being unable to classify a post, 
but in most cases the task was rather straightforward.

Due to extremely time consuming nature of the task, we have chosen 20 threads from the whole set of 58 full discussions. They contained between   20 and 250  posts.

Table~\ref{TabAttrib}  presents percentages of various types of comments between the users (i.e. omitting the classification of comments addressed to the source messages). The reason for this omission was to decouple our analysis from the individual responses to the news article, which has served mostly to distinguish the nodeclass value. Aggressive posts (Dis, Prv, Inv) took almost 75\% of all communications between users -- and we should remember that provocative posts directed to the source news story also add to the `discussion temperature'. 
Agreements between factions were extremely rare, and are only slightly more frequent between those users with declared opinion (nodeclass A,B) and the unidentified ones (NA nodeclass). 

\begin{table}
	\begin{tabular}{rcccc}
	           & Intra-faction & Inter-faction & Factions-NA & Intra-NA \\
	           &  (A-A,B-B) &  (A-B) & (A-NA, B-NA) &  (NA-NA)  \\	\hline
	      Agr  &     16.9\% &      0.6\% &      2.5\% &      0.6\% \\
	
	      Dis  &      2.1\% &     32.8\% &     11.1\% &      1.2\% \\
	
	      Inv  &      0.2\% &     17.1\% &      3.1\% &      0.7\% \\
	
	      Prv  &      1.1\% &      2.7\% &      1.9\% &      0.2\% \\
	
	      Neu  &      1.1\% &      0.8\% &      1.7\% &      0.2\% \\
	
	      Jst  &      0.4\% &      0.0\% &      0.4\% &      0.1\% \\
	
	      Swi  &      0.1\% &      0.0\% &      0.2\% &      0.1\% \\
	\end{tabular}
	\caption{Statistics of comment type between various groups of users (two identified factions A and B and neutral or unidentifiable class NA). 
	\label{TabAttrib} }
\end{table}

From detailed analysis of the discussions it is clear that sub-threads related to neutral posts usually died much faster than those due to confrontational or abusive ones.
Long chains of invective-invective and invective-disagreement are very frequent, while exchanges between users of the same group agreeing with each other are usually shorter, rarely extending beyond 3-4 consecutive posts.

\subsection{Factual and emotional content considerations}

\begin{figure*}
\includegraphics[scale=.9]{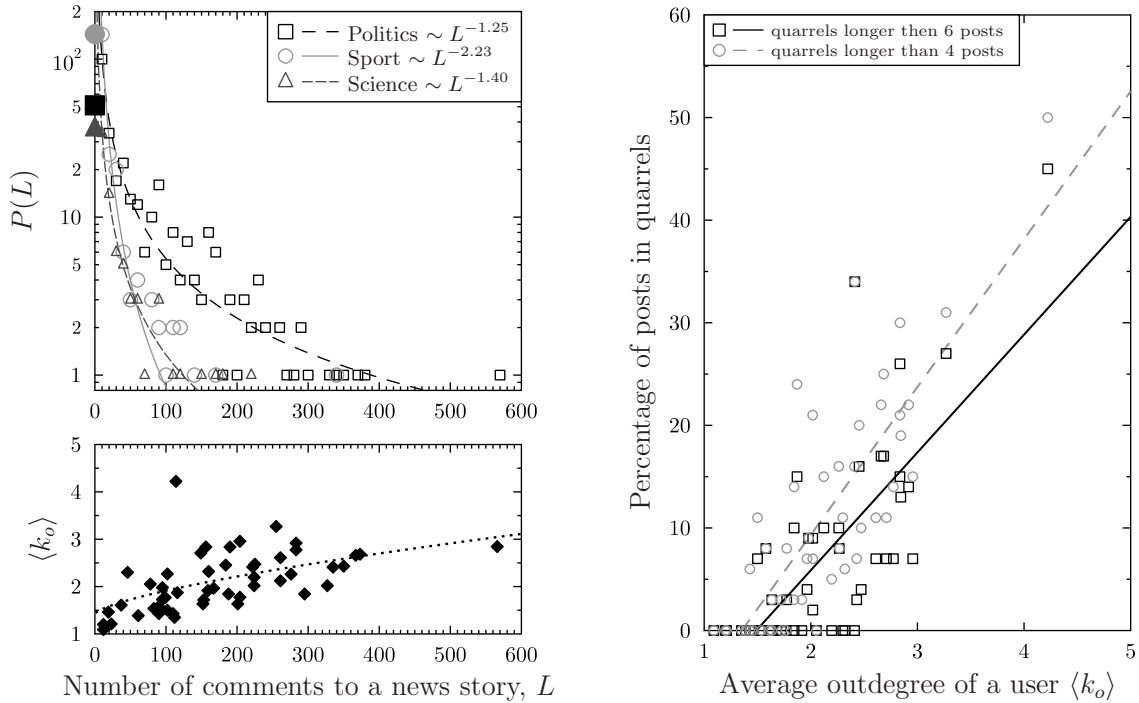} 
\caption{Top left panel: Distribution $P(L)$ of discussions sizes $L$ for three forum topics: \emph{politics}, \emph{sport} and \emph{science}.
Filled points show the number of news items that did not elicit any comments. Lines show power law fits.
The large difference of exponent for the sport forum is due to much larger number of news items, most of which get no or almost no reaction at all. 
Bottom left panel: average outdegree $\left< k_o \right>$, for the \emph{politics} forum as function of $L$. Right panel: correlation between the $\left< k_o \right>$ and percentage of discussion spent on binary exchanges (quarrels). Points show percentages of quarrels longer than 6 posts and longer than 4 posts. The data support supposition that large $\left< k_o \right>$ values are due to extended quarrels between a few participants. \label{FigThreadSizes} 
}
\end{figure*}

To see if these features are indeed characteristic for the strongly polarized political forum, 
we have gathered similar data for two other topics: \emph{sport} and \emph{science}. One would expect comparable 
level of emotions to be present for a sport column, while much lower level for science related news.
As we can see from Fig.~\ref{FigThreadSizes}, in all three cases there is a main group 
of discussions, with size distribution $P(L)$ falling roughly in power law, but there are discussions with 
very high number of posts $L$. We are interested in the possible origins of such threads.

We note first that to our surprise, the sport forum has shown much faster fall with increasing thread size $L$, 
as well as much higher proportion of news items that do not attract any comment. 
This is most likely due to the fact that sports reporting consists of many items covering all disciplines 
(from soccer through tennis to NBA) and the interests of readers would be distributed over the disciplines. 
During the time we have been gathering our data a few major sports events took place (e.g Australian Open 
tournament, Handball World Championships), in which Polish participants were expected to be successful, 
and where high emotional reactions were present.
Indeed, these were the news stories that resulted in high user participation, 
equalling in size those of the political forum. 
However, the topology of sport discussion networks of comments was different from the political ones. 
For example, the largest discussion, shortly after a dramatic win by the 
Polish team in the handball championships, 
involved 249 participants and 336 posts. But 203 users have posted only a single comment, 
attached to the source message and expressing their joy. 
The longest exchange involved just four posts. 
Upon reflection, the sport forum provides almost perfect contrast to the politics: 
it is not dominated by two factions and the moods and opinions of participants are usually in sync.

The main topic of our study, political comments with high degree of conflict, show a very different behaviour.
There are significantly fewer news stories without any comment. 
Most of the stories with small response consist of weakly connected posts, 
 which relate directly to the news source. But, as the size grows, the proportion of exchanges, 
 duels and provocative output by trolls increases strongly. 
 What is important, the proportion of disagreements is very high and the aggressive behaviour 
 (abuse and provocations) increases even more strongly with the size of the discussion. 
 Only in this forum we found large proportion of mid-sized ($L>100$) discussions, 
 fuelled by a mixture random, preferentially attached comments and  quarrels.  

Compared to sport  discussions on science  were at the other extreme. 
For the most part they are very short and rather dull, with practically no network structure. 
But, from time to time, a strictly scientific topic is translated by the users to one of highly 
loaded subjects. Such was the case of pre-natal research which was discussed from religious/ethical 
point of view. Story about details of last Ice Age was discussed with all the emotions due to current 
environmental issues. In such cases we have observed an even stronger dominance of binary exchanges 
in the general topology. First of all, the number of participants was usually much smaller 
than the number of posts, In one case only 12 people `produced' 215 posts, with 8 of 
them responsible for 208 comments. Binary exchanges longer than 8 consecutive posts took 
72\% of the discussion. It should be noted that while the users disagreed with each other, 
only a few comments were abusive, 
and many used evidence, references and logical arguments. 
In another, smaller thread, a discussion between two participants consisted of 40 posts 
(out of a total of 117), while in yet another, two users have exchanged 41 posts (out of 178), 
most of them rather long and `scientific' in spirit. 
Thus, we propose that when the topic of a science news is received 
by the readers as related to important worldview issue, the 
chance for localized conflict of pairs of users is very high. 
This is coupled with a natural barrier of accessibility 
(many participants in political discussions simply do not 
look into the science section at all), so that the proportion of the 
users capable of adding rational arguments to the discussion is higher.

\subsection{Computer model and simulations}
\label{SecSimul}

To get more detailed insight and control over on the relative importance of the processes described above we have constructed a simplified computer model of the community and discussion process. We have tried to keep the model as simple as possible, to focus on the crucial aspects of the process. In the model discussion participants are simulated by agents with the following characteristics: nodeclass (we have kept only A and B classes, no unknowns or neutrals, although they are within the program capabilities) and activity class (high or low level of activity). 

The simulation process is rather typical: agents are selected randomly, 
and then `choose' the post they wish to comment on, from the set of earlier posts. 
Certain proportion of the agents look directly to the `source' message. For others
we use preferential attachment rules for probability of picking the target post. 
Specifically, the chance of choosing a post is proportional to its total degree 
(outdegree of a post is always 1, indegree may be quite high).
After choosing the target post, the agent then decides whether or not to comment on it. 
The probability depends on the activity class and on the nodeclass of the author of target remark. 
In the simplified model, agents of the same class would always agree with each other, 
while agents of different nodeclass would always disagree. 
To describe the  tendency favouring negative responses, 
we have assumed that posting disagreeing comments is more probable 
by a factor of 2 than simple agreements. 
Active participants have higher probability of responding than passive ones. 
Simulation is run until preassigned number of posts are placed, 
at which time suitable statistics are measured.

\begin{figure*}
\includegraphics[scale=0.75]{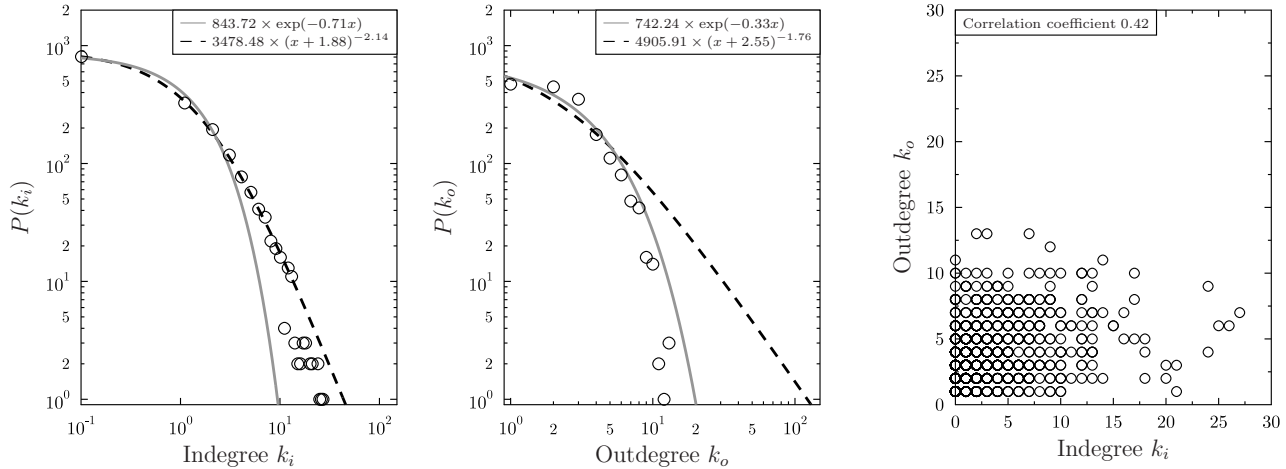} 
\caption{Indegree and outdegree distributions obtained from computer simulations without quarrels. The third panel shows poor correlation between $k_i$ and $k_o$. \label{FigSimulativeNoQuarrels}  
% perhaps add similar statistics for sport, art science?
}
\end{figure*}

Results of such model are presented  Fig.~\ref{FigSimulativeNoQuarrels}. 
The indegree distribution shows some similarity with observations and good 
agreement with power-law distribution, expected for preferential attachment rules. 
On the other hand, outdegree shows significantly faster, exponential decrease rather than power-law. 
This  is not surprising, as it corresponds 
to probabilistic choice of agents posting the comments. 
There are no agents with unusually high values of indegree and outdegree, 
nor is there any significant correlation between $k_i$ and $k_o$.
We conclude that the model needs modification to be able to describe our observations.

The key enhancement of the model is  an additional step in the simulation process. 
Specifically, after the agent has posted a comment, the `author' 
of the target post is `given a chance' to respond. 
The probability of such response is assumed higher than for a normal post 
(to reflect the situation when an agent might `feel personally interested' in responding).
If the response is placed, the roles of the agents are reversed, and again a chance for counter-response is evaluated. 
This chain is continued until one of the agents `decides' to quit. 
The exchange between the two is decoupled from the rest of simulation and it is possible to derive simple analytical 
formulae for the mean length of such exchange. 
Proper choice of response probabilities can be achieved by comparing the lengths of simulated and observed quarrels, 
and is partially independent measure of agreement between simulation and reality.

\begin{figure*}
\includegraphics[scale=0.75]{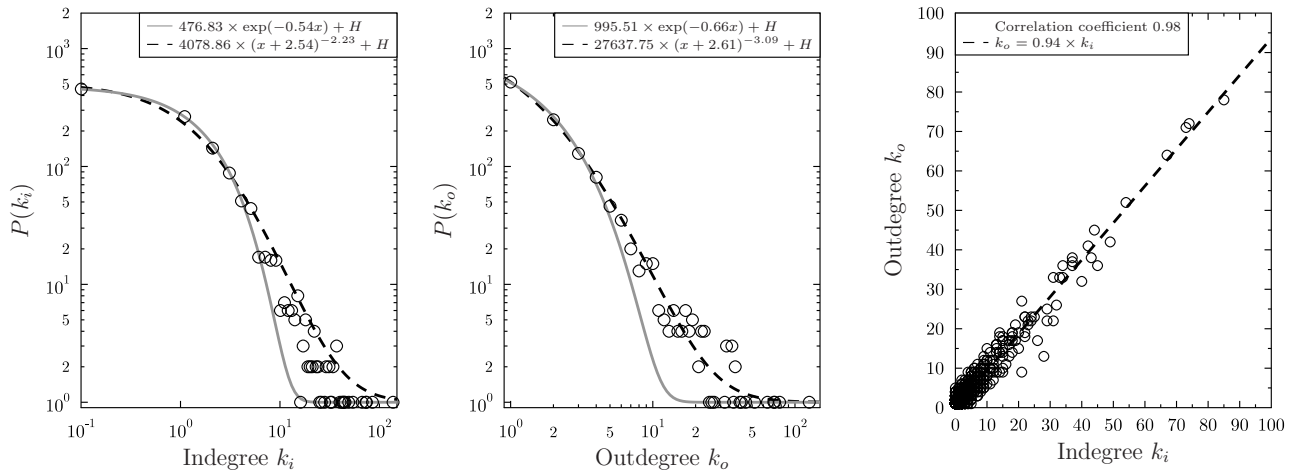} 
\caption{Indegree and outdegree distributions obtained from computer simulations. The third panel shows correlation between $k_i$ and $k_o$. \label{FigSimulativeQuarrels}  
% perhaps add similar statistics for sport, art science?
}
\end{figure*}

When the  quarrels are added into the simulations, the results  become much closer to the observations, 
as shown in Fig.~\ref{FigSimulativeQuarrels}. The similarity  goes beyond the indegree 
and outdegree distributions and extends to details of the network structure,,
for example for the correlation coefficient between $k_i$ and $k_o$. 
Thus, even though radically simplified (no neutral posts, fully polarized opinions of agents, 
simple process), the model yields results surprisingly close to reality.

We note that the computer model described above is able to 
reproduce the characteristic results for the three mentioned forums, 
by simple adjustments of probabilities of posting and entering into duel.

One of the reasons for the success of the simulations  might be the fact that in  
large part of the analysed discussions \textbf{people do behave like mindless automata}. 
Many comments are almost automatic responses to abuse -- in form of further abuse.
Others are canned accusations of the supporters of the opposing political 
faction for being liars, thieves, idiots or worse.
There are very few explanatory or evidence posts (unlike in topical forums where many users aim at helping each other). Even if such evidence appears, it is immediately, almost automatically questioned, as coming `from the other side'. In fact, after reading so many posts we could with high probability predict the character of the post by looking up the names of the participants, without reading the discussion.

\section{Discussion}

\subsection{Citation networks}

One of the topics where network approach has a long history of successful use  is 
the analysis of scientific citations. On can refer here to 
\citet{newman00-2,newman01-2,newman01-4,newman04-1},  
\citet{redner98-1,silagadze99-1,gupta01-1} and \citet{vazquez01-1}.

There are a lot of functional similarities between the internet forum discussions and 
research publication citations. In both cases the participants share some common interest,
although not necessarily the viewpoint or opinion. In both cases communication is indirect, 
via links to messages or publications, rather than directly between people. Also, the lifetime of 
communication is limited. Of course the time-scales are vastly different (hours compared to years)
but this is mitigated by the ease of posting a comment and resulting speed of reaction, 
compared to rather slow scientific publication process.

In addition to these similarities, there are, however, serious differences. 
Probably the most important are due to attitude. Science is a work of cooperation and 
even if authors hold opposing opinions on some topic, the publications usually 
reflect the common aim of getting closer to a true explanation.
While it would be untrue to claim that there is no  personal conflict in science 
(in private or even in public),  there is little room for personal animosity in refereed publications.
This lowers the level of emotional reactions. 
Moreover, there is no anonymity in scientific publications and due to collaborative nature of 
modern research the authors themselves form a complex network of interdependence.
Lastly, the content of scientific publications is dominated by evidence and evaluation, 
and as depersonalised as possible.

In summary, we have two social phenomena which share some of the \emph{technical} 
aspects of indirect, mediated communication, while differing strongly in psychological 
base and content. It is quite interesting to look for similarities and differences
in the network structures of the two cases.

Scientific citation networks have been shown to exhibit modified power law statistics, especially for the large $k_i$ part of the distribution of citations, for example \citet{newman04-2}.
There are, however, also models using different forms of distribution. 
For example \citet{redner98-1} has initially postulated that the indegree of citation
network is well modelled by stretched exponential $P(k_{i})\sim\exp(-\beta k_{i}^{\gamma})$
for low $k_i$ values. The existence of hugely popular papers has been given various explanations.
Some refer to content and importance, pointing out the correlation between citation statistics and 
real importance of the paper or the author. There are also explanations which rely only on statistical properties,
such as  a very simple computer model of `randomly citing scientist', 
which has reproduced remarkably closely the actual distribution of frequency of citing 
(\citet{simkin03-3,simkin04-1,simkin03-1,simkin06-1}).

The general shape of indegree and outdegree distributions for citation networks and for internet 
discussions are remarkably close. The main difference is much more pronounced role of 
a few  highly connected users in our case, which we attribute to quarrelling individuals -- 
phenomenon absent in scientific publications, where such exchanges are relatively rare and 
procedurally limited to a single remark/response cycle. It is interesting to observe that 
the social network structure seems rather stable, regardless of whether the connections 
are motivated by common interests or by spite and hate.

A very interesting analysis of the first mover advantage in scientific
publications has appeared recently (\citet{newman08-1}). 
He has shown that there is significant bias promoting
citations to early papers in a distinct field of research, suggesting, tongue-in-cheek, 
that it is a better strategy to write the first paper in the field than to write the best one.
This bias is, in a sense, related to publication and citation mechanisms, not to actual content.
But Newman's points out that even where the first-mover effect is strong, a small number of later papers
attract significant attention in defiance of advantage of the earlier ones.

Despite the fact that the motivation for posting is radically different, the same phenomenon is observed
in the network discussions. The reason is again technical. Most of the heated exchanges are related with early posts. This is due to 
the way the discussion is visually fragmented into pages containing 100 posts -- viewing the later comments 
requires more effort. Thus the late responses to the original news story 
 are not immediately visible and at a disadvantage compared to early ones. 
Yet, if there is an interesting discussion, the reader can follow it by pressing the `next' button, 
skipping the boundaries between cumulative displays. 
In such situation some posts might get high response rate despite their late timing.

\subsection{Internet networks, blogs and hate groups}

Similar comparison may be attempted between the comment networks and the 
ones formed by Internet web sites.
Here again we find some resemblance and some differences. The most important differences are that web 
sites and their links are much more stable than comments -- 
usually more thought is given by the authors 
when deciding what their pages should be linked to. Also, these links are usually driven by common 
interest and views. One seldom finds links to web sites showing opposite viewpoint -- 
web pages offer a good example of a system where `birds of a feather group together'. 
The last difference might be that generally there is less 
emotion and more content in traditional web pages. 

As in the case of citation networks, despite the differences mentioned above, 
the general statistical properties of discussion threads and web site links are remarkably universal.
For example the modified power-law distribution of page degree is very similar to our case.
The discussions share even more common features with \emph{blogs}, where the personal and emotional content is dominant. A study of blogging behaviour in strongly polarized environment of US 2004 Elections has been published by \citet{adamic05-1}. 
It has shown a lot of preference for strong links between blogs of the same political orientation -- the links between opposing blogs were present, but not numerous, limited to about 15\% of the total number of links. This stands in contract to our observations. An explanation that is plausible relies again on the difference between the puprose of the two types of communication. Blogs, especially election campaign ones, are written with the main aim of promoting particular party or candidate. The conflict with the opponents, if present, is secondary. On the other hand, the large percentage of abuse and invectives in the discussion posts whows that the main purpose is to vent the emotions and possibly to incite hate. Thus the percentage of cross-group links is much higher.

Another detailed study of  blogging behaviour by \citet{leskovic07-1} shows similar power law behaviour  for the indegree and outdegree distribution, albeit with different exponent values. An important difference between the two systems is the lack of observed significant correlation between indegree and outdegree for the blogs, with $k_i$ and $k_o$ correlation coefficient of only $0.16$, much lower than 0.85 in our networks dominated by bilateral exchanges. \citet{leskovic07-1} propose a cascading model of blog links and provide data on relative probability of various patterns of link connections. The binary exchanges of our approach which would correspond to linear topology of the cascade model are relatively less probable in the blog case, where the cascades tend to be 
wide and not deep.

%---
The discussions studied in this work are by no means the only examples of hate present in the vast space of the Internet. 
There are already works which study the network structure of `Hate groups'
(\citet{chau06-1,chau06-2}). These studies are important for us for two reasons. 
First, they focus on  bloggers, who enjoy a lot of freedom to express their opinions and emotions. 
Second, the authors use networking methods, similar to the ones employed here. 
The network of users is formed
through formal subscriptions between blogs and through impromptu comments posted to each other. 
This last aspect corresponds directly to our situation. 
While the political views studied by Chau and Xu are probably more extreme than 
the ones of the readers of the \texttt{www.gazeta.pl} portal, 
the emotional reactions seem to be as strong.
It is quite interesting that the degree distribution for the 
giant component of 273 nodes in Chau and Xu network exhibits 
power-law behaviour, $P(k)\sim k^{-\alpha}$ with exponent $\alpha \approx 1.38$. 
% expand and enhance

\subsection{Implications for consensus formation modelling}

The last conclusion form our observation relates to a different domain.
One of fast growing fields in sociophysics is the study of opinion
formation (for recent review see \citet{castellano07-1}). Most models
use so called `agent based societies' and assume that consensus in
a society forms through a series of exchanges between agents. Depending
on the model, the initial opinions of the agents are changed as a
a result of the interactions. Some models postulate a form of averaging
of opinions towards a mean value (for example \citet{deffuant00-1,hegelsmann02-1}),
others use assumption of one of the interacting agents switching his or her opinion
to fit the others (\citet{sznajd00-1}). 
Consensus formation models have become very popular, unfortunately
in large part the studies concentrate of mathematical formalisms or
Monte Carlo simulations, and not on descriptions of real-life phenomena.
The need of bringing simulations and models closer to reality has
been realized and voiced quite a few times (\citet{moss05-1,epstein08-1,sobkowicz09-1}). 

An interesting result from the present study is that the exchanges
studied here (voicing of opinions in a quasi-anonymous medium) \textbf{do
not lead to consensus formation at all! } If anything, the exchanges
lead to increased rift between the participants. This effect should
be studied in much details, as it possibly bears strongly on the usability
of the models of consensus formation.

On one hand, we could assume that this is a phenomenon specific to
computer mediated interactions, with their lack of face-to-face effects
of increased responsibility, shyness, induced submissiveness and even
sympathy. Anonymity and lack of fear of retribution might embolden
the participants and also promote additional mischievousness (clearly
visible in the presence of provocative posts). Thus one might assume
that the studied form of exchanges is an exception to the general
rules of opinions getting closer as result of interactions.

But everyday experience shows that also when people meet face-to-face,
with full use of  non-verbal and emotional
communication, the clash of opinions seldom leads to an agreement
or even weakening of differences. Both  history and literature
are full of examples of undying feuds, where acts of aggression follow each other, 
from Shakespearean Verona families
to modern political or ethnic strife. Encounters and links between
holders of conflicting opinions often lead to strengthening of opposing
convictions -- in some cases additionally leading to severing the links
between individuals and groups. 

The observations of the Internet discussions should therefore be augmented
by sociological data on flesh-and-blood conflicts and arguments, and
the dynamics of the opinion shifts. But even before such studies are
done or referred to (which the present authors feel is beyond their
competence) the basic assumptions of the sociophysical modelling of
consensus formation should be reworked. This is a very interesting
task, because ostensibly we are faced with two incompatible sets of
observations:

\begin{itemize}
\item Even in the face of repeated information exchanges, with some of them
using `hard data' and evidence to support their viewpoint, participants
in the Internet discussions tend to hold to their opinions, perhaps
even strengthening their resolve with each exchange. Within the analysed
subset of the discussions the conversion of opinion -- even a
simple agreement to a statement from opposing side -- was virtually
absent. Interactions do not seem to lead to opinion averaging or switching.
\item Yet, most of the participants do have well defined opinions. These
must have formed in some way. There are studies indicating
genetic/biological base for some of the political tendencies 
(\citet{jost03-2,alford05-1,amodio07-1,haidt07-1}).
So perhaps the participants in our discussions did have a built-in tendency to pick one
of the sides of the divide, and to stick to it. 
Regardless of genetic considerations the  political attitudes are 
thought to be dependant on fairly stable elements, such as childhood environment,
which again decreases the chances of reaching a consensus.
On the other hand the opinions on
specific events, laws or people could not be neither genetically coded, not due
to cultural formation -- they 
must be reached quickly. Disputants, at one stage or the other become
convinced that Mr X. is a hero -- or a villain. Which means that there
must be some effective mechanisms of opinion formation. 
Certainly, the Internet discussions do not provide such mechanism. 
The question
is -- how to describe such process in a `sociophysical' framework, so far strongly preoccupied with 
a model that assumes that encounters can and do change our attitudes?
\end{itemize}

The persistence of differences of opinions exhibited in online discussions
studied in this work stands in contrast to observations of \citet{wu08-1,wu08-2},
who measured a strong tendency towards moderate views in the course
of time for book ratings posted on \texttt{Amazon.com}. However,
there are significant differences between book ratings and expression
of political views. In the first case the comments are generally favourable
and the voiced opinions are not influenced by personal feuds with
other commentators. Moreover, the \emph{spirit} of book review
is a positive one, with the official aim of providing useful information
for other users. This helpfulness of each of the reviews is measured and displayed,
which promotes pro-sociality and `good' behaviour. In the case of political
disputes it is often the reception in one's own community that counts, 
the show of force and verbal bashing of the opponents. The goal of being admired by 
supporters and hated by opponents promotes very different actions than in the cooperative activities. 
For this reason, there is little to be gained by a commentator when
placing moderate, well reasoned posts -- neither the popularity nor
status is increased. 

Our results suggest possible future models of consensus formation that would take 
into account not only factors leading to convergence of opinions, but also their divergence. 
Nonlinear interplay between these tendencies might lead to interesting results, 
perhaps closer to actual social situations than it is today.

%\bibliographystyle{plainnat}
%\bibliography{../Bibliography/JetBibliography}

\begin{thebibliography}{41}
\expandafter\ifx\csname natexlab\endcsname\relax\def\natexlab#1{#1}\fi
\expandafter\ifx\csname bibnamefont\endcsname\relax
  \def\bibnamefont#1{#1}\fi
\expandafter\ifx\csname bibfnamefont\endcsname\relax
  \def\bibfnamefont#1{#1}\fi
\expandafter\ifx\csname citenamefont\endcsname\relax
  \def\citenamefont#1{#1}\fi
\expandafter\ifx\csname url\endcsname\relax
  \def\url#1{\texttt{#1}}\fi
\expandafter\ifx\csname urlprefix\endcsname\relax\def\urlprefix{URL }\fi
\providecommand{\bibinfo}[2]{#2}
\providecommand{\eprint}[2][]{\url{#2}}

\bibitem[{\citenamefont{Jeong}(2003)}]{jeong03-1}
\bibinfo{author}{\bibfnamefont{A.~C.} \bibnamefont{Jeong}},
  \bibinfo{journal}{The American Journal of Distance Education}
  \textbf{\bibinfo{volume}{17}}, \bibinfo{pages}{25} (\bibinfo{year}{2003}).

\bibitem[{\citenamefont{Jeong}(2005)}]{jeong05-1}
\bibinfo{author}{\bibfnamefont{A.~C.} \bibnamefont{Jeong}},
  \bibinfo{journal}{Distance Education} \textbf{\bibinfo{volume}{26}},
  \bibinfo{pages}{367} (\bibinfo{year}{2005}).

\bibitem[{\citenamefont{Dorogovtsev and Mendes}(2003)}]{dorogovtsev03-2}
\bibinfo{author}{\bibfnamefont{S.~N.} \bibnamefont{Dorogovtsev}}
  \bibnamefont{and} \bibinfo{author}{\bibfnamefont{J.~F.~F.}
  \bibnamefont{Mendes}}, \emph{\bibinfo{title}{Evolution of {N}etworks {F}rom
  {B}iological {N}ets to the {I}nternet and {WWW}}} (\bibinfo{publisher}{Oxford
  University Press}, \bibinfo{year}{2003}).

\bibitem[{\citenamefont{Albert and Barab{\'a}si}(2002)}]{albert02-1}
\bibinfo{author}{\bibfnamefont{R.}~\bibnamefont{Albert}} \bibnamefont{and}
  \bibinfo{author}{\bibfnamefont{A.~L.} \bibnamefont{Barab{\'a}si}},
  \bibinfo{journal}{Review of {M}odern {P}hysics}
  \textbf{\bibinfo{volume}{74}}, \bibinfo{pages}{67} (\bibinfo{year}{2002}).

\bibitem[{\citenamefont{Newman}(2000)}]{newman00-1}
\bibinfo{author}{\bibfnamefont{M.~E.~J.} \bibnamefont{Newman}},
  \bibinfo{journal}{J. {S}tat. {P}hys.} \textbf{\bibinfo{volume}{101}},
  \bibinfo{pages}{819} (\bibinfo{year}{2000}).

\bibitem[{\citenamefont{Newman et~al.}(2002)\citenamefont{Newman, Watts, and
  Strogatz}}]{newman02-1}
\bibinfo{author}{\bibfnamefont{M.~E.~J.} \bibnamefont{Newman}},
  \bibinfo{author}{\bibfnamefont{D.~J.} \bibnamefont{Watts}}, \bibnamefont{and}
  \bibinfo{author}{\bibfnamefont{S.~H.} \bibnamefont{Strogatz}},
  \bibinfo{journal}{Proc. {N}atl. {A}cad. {S}ci. {USA}}
  \textbf{\bibinfo{volume}{99}}, \bibinfo{pages}{2566} (\bibinfo{year}{2002}).

\bibitem[{\citenamefont{Adar}(2006)}]{adar06-1}
\bibinfo{author}{\bibfnamefont{E.}~\bibnamefont{Adar}}, in
  \emph{\bibinfo{booktitle}{CHI}} (\bibinfo{year}{2006}),
  \urlprefix\url{http://graphexploration.cond.org/chi2006/guess-chi2006.pdf}.

\bibitem[{\citenamefont{Adar and Miryung}(2007)}]{adar07-1}
\bibinfo{author}{\bibfnamefont{E.}~\bibnamefont{Adar}} \bibnamefont{and}
  \bibinfo{author}{\bibfnamefont{K.}~\bibnamefont{Miryung}}, in
  \emph{\bibinfo{booktitle}{ICSE}} (\bibinfo{year}{2007}),
  \urlprefix\url{http://cond.org/icse2007-final.pdf}.

\bibitem[{\citenamefont{Newman et~al.}(2001)\citenamefont{Newman, Strogatz, and
  Watts}}]{newman01-1}
\bibinfo{author}{\bibfnamefont{M.~E.~J.} \bibnamefont{Newman}},
  \bibinfo{author}{\bibfnamefont{S.~H.} \bibnamefont{Strogatz}},
  \bibnamefont{and} \bibinfo{author}{\bibfnamefont{D.~J.} \bibnamefont{Watts}},
  \bibinfo{journal}{Phys.{R}ev. {E}} \textbf{\bibinfo{volume}{64}},
  \bibinfo{pages}{026118} (\bibinfo{year}{2001}).

\bibitem[{\citenamefont{Huberman et~al.}(2008)\citenamefont{Huberman, Romero,
  and Wu}}]{huberman08-1}
\bibinfo{author}{\bibfnamefont{B.~A.} \bibnamefont{Huberman}},
  \bibinfo{author}{\bibfnamefont{D.~M.} \bibnamefont{Romero}},
  \bibnamefont{and} \bibinfo{author}{\bibfnamefont{F.}~\bibnamefont{Wu}},
  \emph{\bibinfo{title}{Crowdsourcing, attention and productivity}}
  (\bibinfo{year}{2008}), \urlprefix\url{http://arxiv.org/pdf/0809.3030v1}.

\bibitem[{\citenamefont{Newman}(2004{\natexlab{a}})}]{newman00-2}
\bibinfo{author}{\bibfnamefont{M.~E.~J.} \bibnamefont{Newman}}, in
  \emph{\bibinfo{booktitle}{Complex Networks}}, edited by
  \bibinfo{editor}{\bibfnamefont{E.}~\bibnamefont{Ben-Naim}},
  \bibinfo{editor}{\bibfnamefont{H.}~\bibnamefont{Frauenfelder}},
  \bibnamefont{and} \bibinfo{editor}{\bibfnamefont{Z.}~\bibnamefont{Toroczkai}}
  (\bibinfo{publisher}{Springer, Berlin}, \bibinfo{year}{2004}{\natexlab{a}}),
  vol.~\bibinfo{volume}{64}, pp. \bibinfo{pages}{337--370}.

\bibitem[{\citenamefont{Newman}(2001{\natexlab{a}})}]{newman01-2}
\bibinfo{author}{\bibfnamefont{M.~E.~J.} \bibnamefont{Newman}},
  \bibinfo{journal}{Physical {R}eview {E}} \textbf{\bibinfo{volume}{64}},
  \bibinfo{pages}{016131} (\bibinfo{year}{2001}{\natexlab{a}}).

\bibitem[{\citenamefont{Newman}(2001{\natexlab{b}})}]{newman01-4}
\bibinfo{author}{\bibfnamefont{M.~E.~J.} \bibnamefont{Newman}},
  \bibinfo{journal}{Proc. {N}atl. {A}cad. {S}ci. {USA}}
  \textbf{\bibinfo{volume}{98}}, \bibinfo{pages}{5955}
  (\bibinfo{year}{2001}{\natexlab{b}}).

\bibitem[{\citenamefont{Newman}(2004{\natexlab{b}})}]{newman04-1}
\bibinfo{author}{\bibfnamefont{M.~E.~J.} \bibnamefont{Newman}},
  \bibinfo{journal}{Proc. {N}atl. {A}cad. {S}ci. {USA}}
  \textbf{\bibinfo{volume}{101}}, \bibinfo{pages}{5200}
  (\bibinfo{year}{2004}{\natexlab{b}}).

\bibitem[{\citenamefont{Redner}(1998)}]{redner98-1}
\bibinfo{author}{\bibfnamefont{S.}~\bibnamefont{Redner}},
  \bibinfo{journal}{Eur. {P}hys. {J}.} \textbf{\bibinfo{volume}{B4}},
  \bibinfo{pages}{131} (\bibinfo{year}{1998}).

\bibitem[{\citenamefont{Silagadze}(1997)}]{silagadze99-1}
\bibinfo{author}{\bibfnamefont{Z.~K.} \bibnamefont{Silagadze}},
  \bibinfo{journal}{Complex {S}yst.} \textbf{\bibinfo{volume}{11}},
  \bibinfo{pages}{487} (\bibinfo{year}{1997}).

\bibitem[{\citenamefont{Gupta et~al.}(2005)\citenamefont{Gupta, Campanha, and
  Pesce}}]{gupta01-1}
\bibinfo{author}{\bibfnamefont{H.}~\bibnamefont{Gupta}},
  \bibinfo{author}{\bibfnamefont{J.}~\bibnamefont{Campanha}}, \bibnamefont{and}
  \bibinfo{author}{\bibfnamefont{R.}~\bibnamefont{Pesce}},
  \bibinfo{journal}{Brazilian Journal of Physics}
  \textbf{\bibinfo{volume}{35}}, \bibinfo{pages}{981} (\bibinfo{year}{2005}).

\bibitem[{\citenamefont{Vazquez}(2001)}]{vazquez01-1}
\bibinfo{author}{\bibfnamefont{A.}~\bibnamefont{Vazquez}},
  \emph{\bibinfo{title}{{S}tatistics of citation networks}},
  \bibinfo{howpublished}{\url{ http://arxiv.org/abs/cond-mat/0105031 }}
  (\bibinfo{year}{2001}).

\bibitem[{\citenamefont{Newman}(2005)}]{newman04-2}
\bibinfo{author}{\bibfnamefont{M.~E.~J.} \bibnamefont{Newman}},
  \bibinfo{journal}{Contemporary Physics} \textbf{\bibinfo{volume}{46}},
  \bibinfo{pages}{323} (\bibinfo{year}{2005}).

\bibitem[{\citenamefont{Simkin and Roychowdhury}(2003)}]{simkin03-3}
\bibinfo{author}{\bibfnamefont{M.~V.} \bibnamefont{Simkin}} \bibnamefont{and}
  \bibinfo{author}{\bibfnamefont{V.~P.} \bibnamefont{Roychowdhury}},
  \bibinfo{journal}{Complex {S}ystems} \textbf{\bibinfo{volume}{14}},
  \bibinfo{pages}{269} (\bibinfo{year}{2003}).

\bibitem[{\citenamefont{Simkin and
  Roychowdhury}(2005{\natexlab{a}})}]{simkin04-1}
\bibinfo{author}{\bibfnamefont{M.~V.} \bibnamefont{Simkin}} \bibnamefont{and}
  \bibinfo{author}{\bibfnamefont{V.~P.} \bibnamefont{Roychowdhury}},
  \bibinfo{journal}{Scientometrics} \textbf{\bibinfo{volume}{62}},
  \bibinfo{pages}{367} (\bibinfo{year}{2005}{\natexlab{a}}).

\bibitem[{\citenamefont{Simkin and
  Roychowdhury}(2005{\natexlab{b}})}]{simkin03-1}
\bibinfo{author}{\bibfnamefont{M.~V.} \bibnamefont{Simkin}} \bibnamefont{and}
  \bibinfo{author}{\bibfnamefont{V.~P.} \bibnamefont{Roychowdhury}},
  \bibinfo{journal}{Annals of {I}mprobable {R}esearch} 
  \bibinfo{pages}{24--27} (\bibinfo{year}{2005}{\natexlab{b}}).

\bibitem[{\citenamefont{Simkin and Roychowdhury}(2006)}]{simkin06-1}
\bibinfo{author}{\bibfnamefont{M.~V.} \bibnamefont{Simkin}} \bibnamefont{and}
  \bibinfo{author}{\bibfnamefont{V.~P.} \bibnamefont{Roychowdhury}},
  \bibinfo{journal}{Significance} \textbf{\bibinfo{volume}{3}},
  \bibinfo{pages}{179} (\bibinfo{year}{2006}).

\bibitem[{\citenamefont{Newman}(2008)}]{newman08-1}
\bibinfo{author}{\bibfnamefont{M.~E.~J.} \bibnamefont{Newman}},
  \emph{\bibinfo{title}{The first-mover advantage in scientific publication}}
  (\bibinfo{year}{2008}), \urlprefix\url{http://arxiv.org/pdf/0809.0522v1}.

\bibitem[{\citenamefont{Adamic and Glance}(2005)}]{adamic05-1}
\bibinfo{author}{\bibfnamefont{L.}~\bibnamefont{Adamic}} \bibnamefont{and}
  \bibinfo{author}{\bibfnamefont{N.}~\bibnamefont{Glance}}, in
  \emph{\bibinfo{booktitle}{Proceedings of the 3rd international workshop on
  Link discovery}} (\bibinfo{year}{2005}), pp. \bibinfo{pages}{36--43}.

\bibitem[{\citenamefont{Leskovec et~al.}(2007)\citenamefont{Leskovec, McGlohon,
  Faloutsos, Glance, and Hurst}}]{leskovic07-1}
\bibinfo{author}{\bibfnamefont{J.}~\bibnamefont{Leskovec}},
  \bibinfo{author}{\bibfnamefont{M.}~\bibnamefont{McGlohon}},
  \bibinfo{author}{\bibfnamefont{C.}~\bibnamefont{Faloutsos}},
  \bibinfo{author}{\bibfnamefont{N.}~\bibnamefont{Glance}}, \bibnamefont{and}
  \bibinfo{author}{\bibfnamefont{M.}~\bibnamefont{Hurst}}, in
  \emph{\bibinfo{booktitle}{SIAM International Conference on Data Mining (SDM
  2007)}} (\bibinfo{year}{2007}).

\bibitem[{\citenamefont{Chau and Xu}(2007)}]{chau06-1}
\bibinfo{author}{\bibfnamefont{M.}~\bibnamefont{Chau}} \bibnamefont{and}
  \bibinfo{author}{\bibfnamefont{J.}~\bibnamefont{Xu}},
  \bibinfo{journal}{International Journal of Human-Computer Studies}
  \textbf{\bibinfo{volume}{65}}, \bibinfo{pages}{57} (\bibinfo{year}{2007}).

\bibitem[{\citenamefont{Chau et~al.}(2006)\citenamefont{Chau, Pokfulam, and
  Xu}}]{chau06-2}
\bibinfo{author}{\bibfnamefont{M.}~\bibnamefont{Chau}},
  \bibinfo{author}{\bibfnamefont{H.}~\bibnamefont{Pokfulam}}, \bibnamefont{and}
  \bibinfo{author}{\bibfnamefont{J.}~\bibnamefont{Xu}}, in
  \emph{\bibinfo{booktitle}{Pacific-Asia Conference on Information Systems,
  Kuala Lumpur, Malaysia. Retrieved May}} (\bibinfo{year}{2006}),
  vol.~\bibinfo{volume}{14}, p. \bibinfo{pages}{2007}.

\bibitem[{\citenamefont{Castellano et~al.}(2009)\citenamefont{Castellano,
  Fortunato, and Loreto}}]{castellano07-1}
\bibinfo{author}{\bibfnamefont{C.}~\bibnamefont{Castellano}},
  \bibinfo{author}{\bibfnamefont{S.}~\bibnamefont{Fortunato}},
  \bibnamefont{and} \bibinfo{author}{\bibfnamefont{V.}~\bibnamefont{Loreto}},
 \bibinfo{journal}{Rev. Mod. Phys.} \textbf{\bibinfo{volume}{81}} \bibinfo{pages}{591--646} (\bibinfo{year}{2009}).

\bibitem[{\citenamefont{Deffuant et~al.}(2000)\citenamefont{Deffuant, Neau,
  Amblard, and Weisbuch}}]{deffuant00-1}
\bibinfo{author}{\bibfnamefont{G.}~\bibnamefont{Deffuant}},
  \bibinfo{author}{\bibfnamefont{D.}~\bibnamefont{Neau}},
  \bibinfo{author}{\bibfnamefont{F.}~\bibnamefont{Amblard}}, \bibnamefont{and}
  \bibinfo{author}{\bibfnamefont{G.}~\bibnamefont{Weisbuch}},
  \bibinfo{journal}{Advances in {C}omplex {S}ystems}
  \textbf{\bibinfo{volume}{3}}, \bibinfo{pages}{87} (\bibinfo{year}{2000}.

\bibitem[{\citenamefont{Hegselmann and Krause}(2002)}]{hegelsmann02-1}
\bibinfo{author}{\bibfnamefont{R.}~\bibnamefont{Hegselmann}} \bibnamefont{and}
  \bibinfo{author}{\bibfnamefont{U.}~\bibnamefont{Krause}},
  \bibinfo{journal}{Journal of Artifical Societies and Social Simulation
 } \textbf{\bibinfo{volume}{5}} (\bibinfo{year}{2002}).

\bibitem[{\citenamefont{Sznajd-Weron and Sznajd}(2000)}]{sznajd00-1}
\bibinfo{author}{\bibfnamefont{K.}~\bibnamefont{Sznajd-Weron}}
  \bibnamefont{and} \bibinfo{author}{\bibfnamefont{J.}~\bibnamefont{Sznajd}},
  \bibinfo{journal}{Int. {J}. {M}od. {P}hys. {C}}
  \textbf{\bibinfo{volume}{11}}, \bibinfo{pages}{1157} (\bibinfo{year}{2000}).

\bibitem[{\citenamefont{Moss and Edmonds}(2005)}]{moss05-1}
\bibinfo{author}{\bibfnamefont{S.}~\bibnamefont{Moss}} \bibnamefont{and}
  \bibinfo{author}{\bibfnamefont{B.}~\bibnamefont{Edmonds}},
  \bibinfo{journal}{Journal of {A}rtificial {S}ocieties and {S}ocial
  {S}imulation} \textbf{\bibinfo{volume}{8}}, \bibinfo{pages}{13}
  (\bibinfo{year}{2005}).

\bibitem[{\citenamefont{Epstein}(2008)}]{epstein08-1}
\bibinfo{author}{\bibfnamefont{J.~M.} \bibnamefont{Epstein}},
  \bibinfo{journal}{Journal of {A}rtificial {S}ocieties and {S}ocial
  {S}imulation}
  \textbf{\bibinfo{volume}{11}}, \bibinfo{pages}{12} (\bibinfo{year}{2008}).

\bibitem[{\citenamefont{Sobkowicz}(2009)}]{sobkowicz09-1}
\bibinfo{author}{\bibfnamefont{P.}~\bibnamefont{Sobkowicz}},
  \bibinfo{journal}{Journal of {A}rtificial {S}ocieties and {S}ocial
  {S}imulation} \textbf{\bibinfo{volume}{12}}, \bibinfo{pages}{11} (\bibinfo{year}{2009}).

\bibitem[{\citenamefont{Jost et~al.}(2003)\citenamefont{Jost, Glaser,
  Kruglanski, and Sulloway}}]{jost03-2}
\bibinfo{author}{\bibfnamefont{J.}~\bibnamefont{Jost}},
  \bibinfo{author}{\bibfnamefont{J.}~\bibnamefont{Glaser}},
  \bibinfo{author}{\bibfnamefont{A.}~\bibnamefont{Kruglanski}},
  \bibnamefont{and} \bibinfo{author}{\bibfnamefont{F.}~\bibnamefont{Sulloway}},
  \bibinfo{journal}{PSYCHOLOGICAL BULLETIN} \textbf{\bibinfo{volume}{129}},
  \bibinfo{pages}{339} (\bibinfo{year}{2003}).

\bibitem[{\citenamefont{Alford et~al.}(2005)\citenamefont{Alford, Funk, and
  Hibbing}}]{alford05-1}
\bibinfo{author}{\bibfnamefont{J.~R.} \bibnamefont{Alford}},
  \bibinfo{author}{\bibfnamefont{C.~L.} \bibnamefont{Funk}}, \bibnamefont{and}
  \bibinfo{author}{\bibfnamefont{J.~R.} \bibnamefont{Hibbing}},
  \bibinfo{journal}{American Political Science Review}
  \textbf{\bibinfo{volume}{99}}, \bibinfo{pages}{153} (\bibinfo{year}{2005}).

\bibitem[{\citenamefont{Amodio et~al.}(2007)\citenamefont{Amodio, Jost, Master,
  and Yee}}]{amodio07-1}
\bibinfo{author}{\bibfnamefont{D.~M.} \bibnamefont{Amodio}},
  \bibinfo{author}{\bibfnamefont{J.~T.} \bibnamefont{Jost}},
  \bibinfo{author}{\bibfnamefont{S.~L.} \bibnamefont{Master}},
  \bibnamefont{and} \bibinfo{author}{\bibfnamefont{C.~M.} \bibnamefont{Yee}},
  \bibinfo{journal}{Nat Neurosci} \textbf{\bibinfo{volume}{10}},
  \bibinfo{pages}{1246} (\bibinfo{year}{2007}).

\bibitem[{\citenamefont{Haidt and Graham}(2007)}]{haidt07-1}
\bibinfo{author}{\bibfnamefont{J.}~\bibnamefont{Haidt}} \bibnamefont{and}
  \bibinfo{author}{\bibfnamefont{J.}~\bibnamefont{Graham}},
  \bibinfo{journal}{Social Justice Research} \textbf{\bibinfo{volume}{20}},
  \bibinfo{pages}{98} (\bibinfo{year}{2007}).

\bibitem[{\citenamefont{Wu and Huberman}(2008{\natexlab{a}})}]{wu08-1}
\bibinfo{author}{\bibfnamefont{F.}~\bibnamefont{Wu}} \bibnamefont{and}
  \bibinfo{author}{\bibfnamefont{B.~A.} \bibnamefont{Huberman}}, in
  \emph{\bibinfo{booktitle}{Proceedings of the Workshop on Internet and Network
  Economics}} (\bibinfo{year}{2008}{\natexlab{a}}).

\bibitem[{\citenamefont{Wu and Huberman}(2008{\natexlab{b}})}]{wu08-2}
\bibinfo{author}{\bibfnamefont{F.}~\bibnamefont{Wu}} \bibnamefont{and}
  \bibinfo{author}{\bibfnamefont{B.~A.} \bibnamefont{Huberman}},
  \emph{\bibinfo{title}{Public discourse in the web does not exhibit group
  polarization}} (\bibinfo{year}{2008}{\natexlab{b}}),
  \urlprefix\url{http://www.hpl.hp.com/research/scl/papers/opinion\_expression%
/discourse.pdf}.

\end{thebibliography}

\end{document}